\documentclass[twocolumn,prb,showpacs,floatfix]{revtex4}
\usepackage{graphicx}
\usepackage{bm}
\usepackage[dvipdfm]{hyperref}
\newcommand{\be}{\begin{equation}}
\newcommand{\ee}{\end{equation}}
\newcommand{\bea}{\begin{eqnarray}}
\newcommand{\eea}{\end{eqnarray}}

\begin{document}

\title{Crossover of spectral features from Dimerized to Haldane
behavior in alternating antiferromagnetic-ferromagnetic spin-half chains}

\author{
         Weihong Zheng$^{(a)}$,
        Chris J.~Hamer$^{(a)}$, and
        Rajiv R.~P.~Singh$^{(b)}$
}
\affiliation{
$^{(a)}$ School of Physics, University of New South Wales, Sydney NSW
2052, Australia\\
$^{(b)}$ Department of Physics, University of California, Davis, CA
95616\\
}

%
%

\date{\today}

\begin{abstract}
We calculate the excitation spectrum
and spectral weights of the alternating antiferromagnetic-ferromagnetic
spin-half Heisenberg chain with exchange couplings $J$ and $-|\lambda|J$
as a power series in $\lambda$.
For small $|\lambda|$, the gapped one-particle spectrum has a maximum
at $k=0$ and there is a rich structure of
bound (and anti-bound) states below (and above) the 2-particle
continuum. As $|\lambda|$ is increased past unity the spectrum crosses over
to the Haldane regime, where the peak shifts away from $k=0$,
the one particle states merge with the bottom of the continuum
near $k=0$, and the spectral weights associated with the one-particle
states become very small. Extrapolation of the spectrum to large $|\lambda|$ confirms
that the
ground state energy and excitation gap map onto those of the spin-one chain.
\end{abstract}

\maketitle


Recently there has been considerable interest in termination
of the spin excitation spectrum in gapped spin-systems,
when a discrete state meets the continuum.
This issue has been studied theoretically\cite{kolezuk,zhitomirsky} and experimentally
in one\cite{ma,masuda,zaliznyak} and higher dimensional\cite{stone} systems. One interesting
observation is that while in the one-dimensional case
the spectrum appears to merge with the bottom of the continuum,\cite{ma}
in the higher dimensional system it appears to enter the
continuum in the form of a broadened resonance.\cite{stone}

Controlled and systematic numerical calculation of the spin dynamics of quantum
spin models remains a challenging computational task.\cite{feiguin}
Despite much progress in developing computational methods
there are few methods that can accurately calculate
even the single-particle spectrum. Calculations of
the multi-particle continuum and bound states remains
even more daunting, with series expansion methods clearly
leading the way.\cite{tre00,zhe01,kne01,knetter}

We have recently developed a linked-cluster formalism to calculate
the single-particle and multi-particle spectra and spectral weights
of quantum spin models
by means of high-order series expansions.\cite{zheng03,hamer03,gel00}
Here, we apply the method to the
alternating antiferromagnetic-ferromagnetic spin-half
Heisenberg chain with Hamiltonian
\begin{equation}
H =  \sum_{i} {\Big [} {\bf S}_{2i}\cdot {\bf S}_{2i+1} +  \lambda {\bf S}_{2i-1}\cdot {\bf S}_{2i}
{\Big ]} \label{eqH}
\end{equation}

This model is particularly interesting in that it
interpolates smoothly between the spin-half chain ($\lambda\to 1$)
and the spin-one chain ($\lambda\to-\infty$). The case of positive $\lambda$
corresponds to the antiferromagnetic alternating Heisenberg chain (AHC)
and has been studied extensively in the literature.\cite{hamer03} Our goal here is to
study the negative $\lambda$ model. In particular we are interested in the
crossover in the spectrum from the dimerized behavior near $\lambda=0$,
where the single-particle dominates the spectrum
and is well separated from the multi-particle continuum,
to the Haldane chain behavior at large $|\lambda|$, where part of
the single-particle spectrum begins to overlap with the two-particle
continuum. One of the interesting questions is, what happens to
the single-particle spectrum and spectral weights as the
single-particle states
meet the two-particle continuum?
Here we use the same notation as in reference \onlinecite{hamer03}.

We also present
quantitative studies of bound and anti-bound states in the model.
The latter can only be done reliably when $|\lambda|$ is not too large,
as it is unclear how series extrapolation methods can be applied
to the multi-particle spectra. We find that there are two
bound/antibound states in each of the $S=0$, $S=1$ and $S=2$ sectors, of which
a bound state
in each sector disappears between $-1>\lambda > -2$.
We also extrapolate the
ground state energy and the spin-gap at $k=\pi$ to $\lambda\to -\infty$
to confirm the mapping of the model to the spin-one Heisenberg chain.\cite{hida}

This model is also interesting from an experimental point of view,\cite{tennant}
as several alternating chains are suspected to have an alternating
ferromagnetic/antiferromagnetic character. One of our motivations is
to present detailed results on spectral weights for bound states
and multi-particle continuua which can help the search for these
features in experiments.

Throughout this paper we assume the inter-dimer spacing is $d$ and
that all spins are equally spaced at a distance $d/2$. We have
calculated the spectrum for a more general geometry, where the
dimers are oriented in some way with respect to the chain, and
where the projected distance between neighboring spins within
the dimer and between dimers may be different.\cite{mikeska} These are likely
very important for a detailed comparison with specific experimental
systems but not important for discussing the general properties.
These series, or the extrapolated plots can be obtained from the
authors on request.

We begin with a discussion of the spectrum for small $|\lambda|$, where
no series extrapolation is needed and simple summation leads to
very accurate results. In Fig.~1, we show the single-particle and
two-particle spectra calculated for the model at $\lambda=-0.5$. One
can see that the one-particle spectrum has a peak at $k=0$ and it
is well separated from the two-particle continuum. Also shown are
the various bound and anti-bound states. It is not difficult to
understand why the dominant bound and antibound states ($S_1$, $T_1$
and $Q_1$) in the singlet, triplet and quintuplet sectors are
reversed with respect to the $\lambda>0$ case. The ferromagnetic
interaction becomes attractive in the $S=2$ channel and repulsive
in the $S=0$ and $S=1$ channels. The behavior of the weaker bound
and antibound states ($S_2$, $T_2$ and $Q_2$) arises from further
neighbor interactions generated in higher orders of perturbation
theory. Hence, they are harder to treat analytically.\cite{zheng03,hamer03}
\begin{figure}[!htb]
\begin{center}
  \includegraphics[width=7.cm,angle=0]{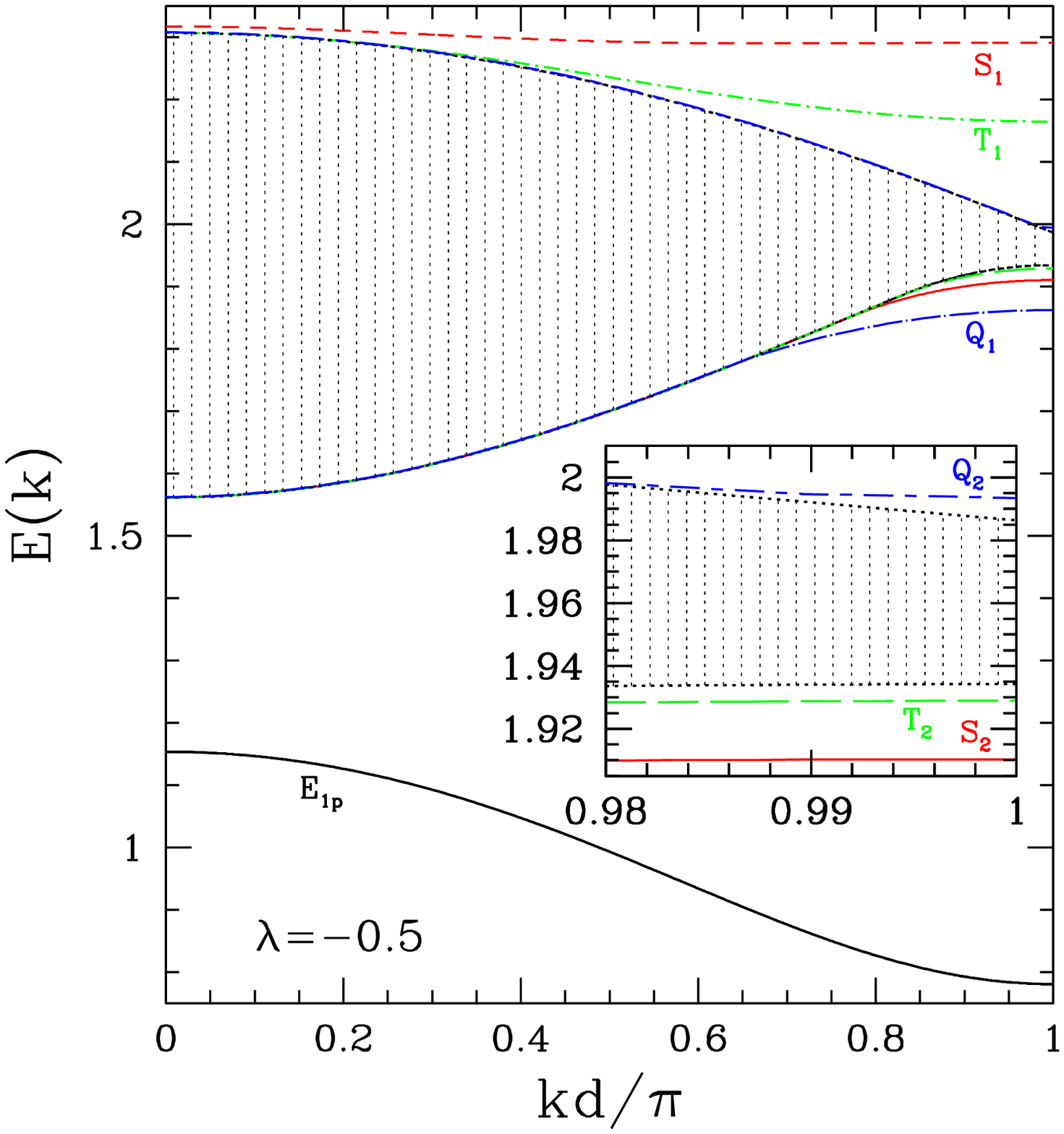}
  \caption{\label{fig1} (Color online) The spectrum at $\lambda=-0.5$. The
one-particle spectrum ($E_{\rm 1p}$) is maximum at $k=0$ and is separated from the
two-particle continuum (gray shaded) by a substantial margin. Also shown is
the two-particle continuum and the three bound ($S_1$, $T_1$
and $Q_2$) and three anti-bound states ($S_2$, $T_2$ and $Q_1$).
The inset shows the bound and antibound states very near $kd=\pi$.}
\end{center}
\end{figure}

In Fig.~2 we show the spectral weights associated with
the various states. It is evident that the one particle state
carries the most weight. For $kd<\pi$ almost all the weight lies
in the single-particle spectrum. The two-particle weights peak
at $kd=2\pi$. In Fig.~3 the relative contributions for one
and two-particle states are shown. At $kd=2\pi$ the two-particle
weights can reach above $13$ percent. The sum of one and two-particle
weights for all $k$ carries about $99.9$ percent of the weight.
Thus, there is negligible spectral weight in states with more than two
particles.
In Fig.~4 the relative weight in
the two bound states is shown as a function of $k$. Note that the {\it
antibound} state $T_1$, lying {\it above} the continuum, still carries the
dominant weight. Both weights remain small
at all $k$ and thus may not be easy to observe experimentally, the
largest weight fraction being less than 4 percent.

\begin{figure}[!htb]
\begin{center}
  \includegraphics[width=6.cm,angle=0]
{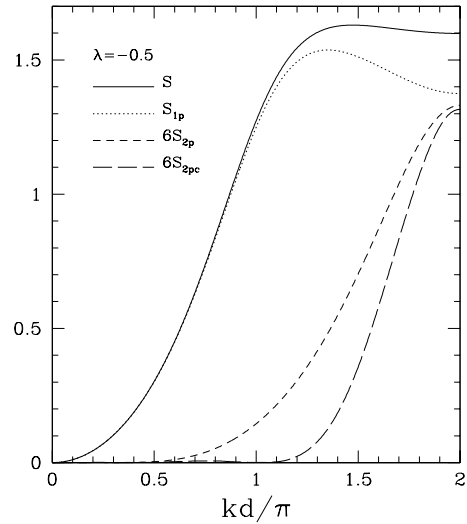}
  \caption{\label{fig2} Integrated structure factor $S$
and spectral weights for
one-particle excitations ($S_{\rm 1p}$), two-particle excitations ($S_{\rm 2p}$),
and the two-particle  continuum ($S_{\rm 2pc}$) at $\lambda=-0.5$.}
\end{center}
\end{figure}
\begin{figure}[!htb]
\begin{center}
  \includegraphics[width=6.5cm,angle=0]
{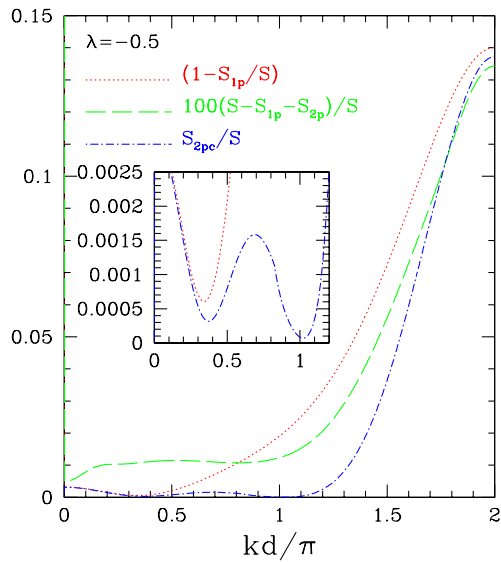}
  \caption{\label{fig3} (Color online) Relative spectral weights for
one and two-particle excitations at different wavevectors for $\lambda=-0.5$.}
\end{center}
\end{figure}

\begin{figure}[!htb]
\begin{center}
  \includegraphics[width=6.5cm,angle=0]
{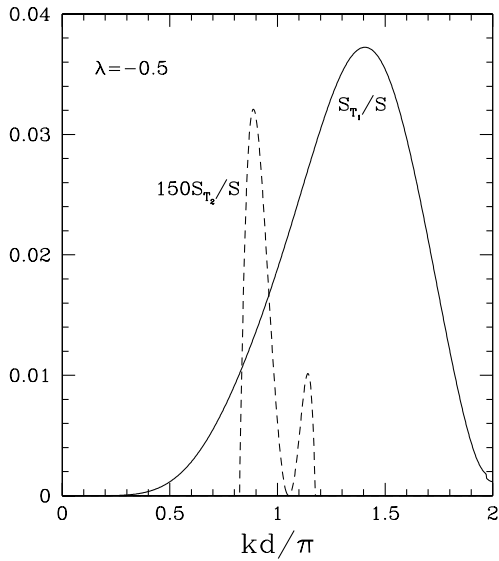}
  \caption{\label{fig4} Relative weights of bound state ($T_2$)
  and antibound state ($T_1$)
at different wavevectors for $\lambda=-0.5$.}
\end{center}
\end{figure}

To see the evolution of the model as $|\lambda|$ is increased past
unity, we need to use series extrapolation methods.\cite{gut} In Fig.~5,
we show the evolution of the integrated structure factor
as a function of $|\lambda|$. We see that the crossover to the
Haldane chain behavior is related to the development of
a short-range antiferromagnetic peak at $kd=\pi$.
Our results at $\lambda=-\infty $ agree with
 an extrapolation of the
 finite lattice results of Takahashi\cite{tak} for the spin-1 chain at $k=\pi$.

\begin{figure}[!htb]
\begin{center}
  \includegraphics[width=6.5cm,angle=0]
{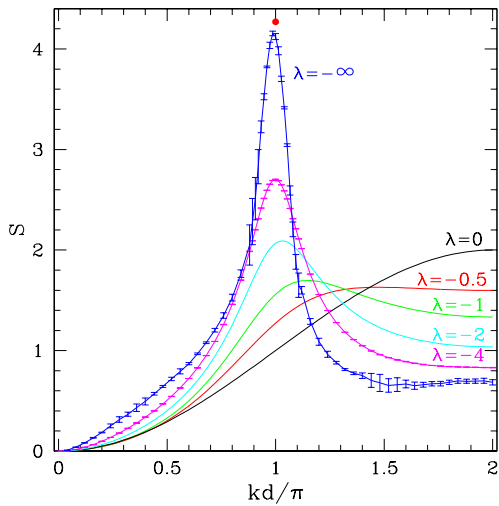}
  \caption{\label{fig5} (Color online) Evolution of the integrated
structure factor $S$ with $|\lambda|$. The red point is an extrapolation of the
 finite lattice results of Takahashi\cite{tak} at $k=\pi$.}
\end{center}
\end{figure}

In Fig. 6, we show the spectrum at $\lambda=-1,-1.5$ and $-2$ obtained through
series extrapolation methods together with the upper and lower boundaries
of the 2-particle continuum. There are several
important features to observe here. First, the peak in the
single-particle spectrum has moved away from $k=0$ and the
spectrum is beginning to resemble more the behavior in
Haldane chains. Second, the spectrum near $k=0$ potentially
overlaps with the two-particle spectrum. We find that the
single-particle spectrum rather than moving into the
continuum and broadening actually merges with the bottom
of the continuum. This is consistent with observations on
the Haldane chain materials.\cite{ma}

\begin{figure}[!htb]
\begin{center}
  \includegraphics[width=6.5cm,angle=0]
{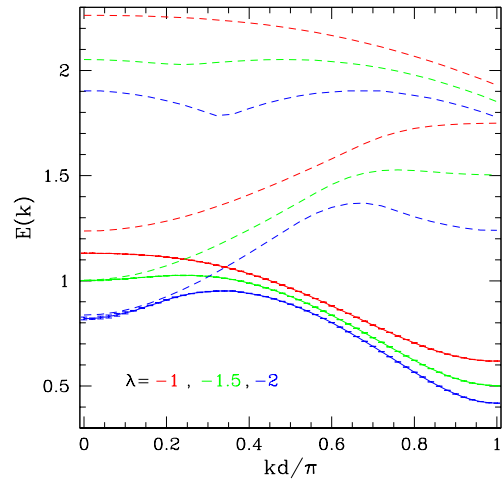}
  \caption{\label{fig6} (Color online) One particle energies (solid curves) and upper
and lower boundaries of the two-particle continuum
(dashed curves)  at different wavevectors for $\lambda=-1.0$ (red curves), $-1.5$ (green curves) and $-2.0$ (blue curves).}
\end{center}
\end{figure}

\begin{figure}[!htb]
\begin{center}
  \includegraphics[width=6.5cm,angle=0]
{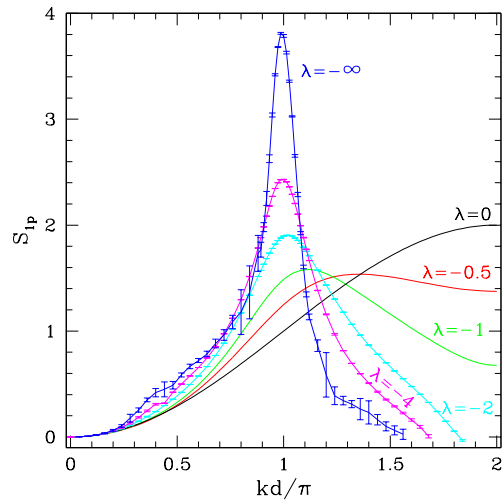}
  \caption{\label{fig7} (Color online) Single-particle spectral
weights ($S_{\rm 1p}$) at different wavevectors as a function of $\lambda$.
}
\end{center}
\end{figure}

In Fig.~7, we show the evolution of
one-particle spectral weights as a function of $\lambda$. We find
that in the region where the single-particle spectrum merges
with the continuum, its spectral weight becomes very small. Note
that the $x$-axis for this figure runs over $0<kd<2\pi$. The
spectra are symmetric around $kd=\pi$ and merge with the bottom
of the continuum near both $kd=0$ and $kd=2\pi$. Near $kd=0$,
the spectral weights are very small to begin with. As the
single-particle states merge with the continuum, the weights also
become very small near $kd=2\pi$.

\begin{figure}[!htb]
\begin{center}
  \includegraphics[width=6.5cm,angle=0]
{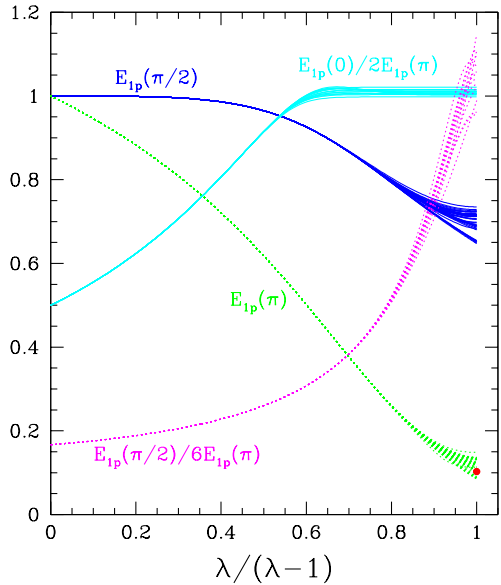}
  \caption{\label{fig8} (Color online) Single-particle energies and
energy ratios at selected wavevectors
as functions of $\lambda$. The red circle indicates the known
gap for the spin-one chain \cite{white}.
Values for various integrated differential approximants are shown.}
\end{center}
\end{figure}

Extrapolating the full spectrum to larger $|\lambda|$ proves difficult.
The spectrum at small $k$ shows poor convergence, which may be related
to the fact that single-particle states are not well defined at small $k$,
and the spectrum is not a monotonically  decreasing function of $\lambda$.
However, extrapolating the ground state energy,
and the single-particle excitation energy in the range $\pi/2<kd<\pi$,
we have verified that the model maps on to the
spin-one chain with an exchange constant of $J/4$, with numerical values
in agreement with DMRG studies on the spin-one chain.\cite{white}
Fig.~8 shows a plot of the estimated single-particle excitation energy
at momenta $kd=\pi/2$ and $kd=\pi$
together with the ratio of those energies and the ratio of the
excitation energy at $k=0$ to that at $kd=\pi$, as functions of
$\lambda$. The latter ratio
saturates at a value of $2$ implying again that the one-particle state merges with the
bottom of the continuum at $k=0$ from this coupling on.
It can be seen that $E_{1p}(\pi)$ maps smoothly onto the energy gap for
the spin-one chain as $\lambda \to -\infty$.

In conclusion, in this paper we have studied the excitation spectra
of the alternating ferromagnetic-antiferromagnetic spin-half chain,
and the crossover from the dimerized phase when the antiferromagnetic
interactions are stronger to the Haldane phase when the ferromagnetic
interactions become stronger. We find that in the former phase the
single-particle states are separated from the two-particle
continuum and there is a rich spectrum of bound states. In the
latter phase the single-particle states are only well defined
over part of the Brillouin zone and merge with the bottom of
the two-particle continuum near $k=0$. We present detailed results
for various spectral weights, including bound states and
continuua, which can be helpful in experimental searches
for these subtle effects.

We thank Prof. Steven Nagler and Dr. Matt Stone, for encouraging us to study
this problem and to Prof. O.P. Sushkov and Prof. J. Oitmaa for discussions.
This work is supported by a grant
from the Australian Research Council and by US National Science Foundation
grant number DMR-0240918.
 We are grateful for the computing resources provided
 by the Australian Partnership for Advanced Computing (APAC)
National Facility.

\bibliography{basename of .bib file}


\end{document}